# Inverse Design of Plasmonic Structures with FDTD

Zhou Zeng and Xianfan Xu*

School of Mechanical Engineering and Birck Nanotechnology Center

Purdue University, West Lafayette, IN 47906

* Corresponding author.

Email addresses: zeng133@purdue.edu (Zhou Zeng), xxu@ecn.purdue.edu (Xianfan Xu)

## Abstract

Inverse design has greatly expanded nanophotonic devices and brought optimized performance. However, the use of inverse design for plasmonic structures has been challenging due to local field concentrations that can lead to errors in gradient calculation when the continuum adjoint method is used. On the other hand, with the discrete adjoint method one can achieve the exact gradient. Historically the discrete version is exclusively used with a Finite Element model, and applying the Finite-Difference Time-Domain (FDTD) method in inverse design of plasmonic structures is rarely attempted. Due to the popularity of using FDTD in simulating plasmonic structures, we integrate the discrete adjoint method with FDTD and present a framework to carry out inverse design of plasmonic structures using density-based topology optimization. We demonstrate the exactness of the gradient calculation for a plasmonic block structure with varying permittivity. Another challenge that is unique with plasmonic structures is that non-physical amplification caused by poorly chosen material interpolation can destroy a stable convergence of the optimization. To avoid this, we adopt a non-linear material interpolation scheme in the FDTD solver. In addition, filtering-and-projection regularization is incorporated to ensure manufacturability of the designed structures. As an example of this framework, successful reconstruction of electric fields of a plasmonic bowtie aperture is presented.

## 1. INTRODUCTION

Plasmonics is one of the most widely studied areas in modern photonics, with a wide range applications including super-resolution imaging,[1,2] nanolithography,[3,4] biosensing[5], and high density data storage.[6,7] Inverse design of plasmonics has been reported in recent years[8–10] and is almost exclusively performed using density-based topology optimization with the Finite Element Method (FEM), where the gradient is calculated using the discrete adjoint method. In this approach, the geometry is divided into small elements of which the material property is parameterized by density parameters and the gradients of the discretized objective function with respect to the density parameters are used to evolve the solution.[11] The other common variation of inverse design[12,13] uses level-set topology optimization with FEM and the continuum adjoint method. Instead of directly associating density parameters with small elements, the level-set approach represents domains and their boundaries as level sets of a continuous function.[14] Also, instead of calculating gradients based on the discretized objective function, the continuum adjoint method calculates gradients based on the original continuum objective function and then evaluates it numerically.[15]

Up till now, topology optimization with the Finite-Difference Time-Domain (FDTD)[16] technique has been rare in inverse design of plasmonic structures. There are two reasons for this. First, using the continuum adjoint method with FDTD is challenging because numerical error in electromagnetic simulations of plasmonic structures is highest at the metal-dielectric interfaces. These locations contribute strongly to design sensitivities[17] and can lead to highly inaccurate sensitivity calculation. By refining conformal boundary representation in the FEM one can obtain highly accurate design sensitivities. However, boundary is not conformal in FDTD and cannot be refined easily either. Second, implementing the discrete adjoint method in FDTD is more difficult than in the FEM because it does not have a direct matrix representation of the system and thus requires more customized programming. As a result, most literatures use FEM in inverse design of plasmonic structures. Due to popularity of using FDTD in simulating plasmonics, we propose a new way to carry out inverse design of plasmonic structures with a FDTD solver. Since plasmonics is mostly concerned with frequency-domain response, results from FDTD are first converted into frequency domain and then analyzed. We then incorporate the frequency-domain discrete adjoint method into a density-based topology optimization framework. This method can be built upon an existing FDTD solver and accepts any differentiable frequency-domain objective function. Another challenge unique with plasmonic structures is that non-physical field amplification can occur in and around spatial regions containing non-physical mixtures of material densities and can destroy the stable convergence of the design procedure towards physically admissible designs.[9] A non-linear material interpolation scheme is demonstrated to improve the behavior of the topology optimization compared to the linear or inverse material interpolation and is implemented in our method.

There are two original methods of sensitivity calculation of plasmonic structures with FDTD that can be used in inverse design. The adjoint variable method (AVM)[18] considers the time marching scheme in FDTD as a dynamic system and uses time traversal to simulate the adjoint system. This method uses the discrete adjoint method and is shown to work reasonably well for both dielectrics[19] and plasmonics.[20] However, the objective function is restricted to a time-domain integral function. It is not applicable to functions that has a non-linear dependence on frequency-domain variables such as the optical absorption. Our method transforms the dynamic system into frequency domain and is compatible with general differentiable frequency-domain functions. The other method is based on topological derivatives[21] and has been applied in design of a near-field transducer for heat-assisted magnetic recording.[22] It evolves a bit-map representation of structure by metalizing bits at the boundary and in the void. Each bit is either metalized or nonmetallized based on changes of the objective function due to perforation of the material domain by an infinitesimal hole. However, it calculates topological derivatives based on the original continuum governing equation and is not immune to errors caused by field concentrations.

The paper is structured as follows: We first describe the derivation of the frequency-domain discrete adjoint method with FDTD and validate its accuracy with a numerical example. Subsequently, we introduce a density-based topology optimization method that incorporates the discrete adjoint method. Filtering-and-projection regularization and the non-linear material interpolation scheme are applied to jointly improve robustness. The method is illustrated through reconstruction of electric fields of a plasmonic bowtie aperture antenna.

# 2. DISCRETE ADJOINT METHOD: FDTD IN FREQUENCY DOMAIN

## 2.1 Derivation of the Adjoint Method

To convert the leapfrog time-stepping scheme of FDTD into a frequency-domain matrix representation, we apply the discrete-time fourier transform (DTFT).[23] We present the derivation in **Supporting Information 1**. In short, the joint operations of FDTD discretization and DTFT convert the original Maxwell's curl equations

$$\begin{pmatrix} -\dfrac{\partial}{\partial t}\varepsilon * & \nabla \times \\ \nabla \times & \dfrac{\partial}{\partial t}\mu * \end{pmatrix}\begin{pmatrix} \mathbf{E} \\ \mathbf{H} \end{pmatrix} = \begin{pmatrix} \mathbf{J} \\ -\mathbf{M} \end{pmatrix} \tag{1}$$

into a frequency-domain matrix representation

$$\begin{pmatrix} -i\omega D_e(z) & \mathbf{C}_h \\ \mathbf{C}_e & i\omega D_h(z) \end{pmatrix}\begin{pmatrix} \tilde{e}(z) \\ \tilde{h}(z) \end{pmatrix} = \begin{pmatrix} \tilde{j}(z) \\ -\widetilde{m}(z) \end{pmatrix} \tag{2}$$

where $z = e^{i\omega\Delta t}$ is used for simplicity and tilde denotes DTFT of the corresponding variable. Column vector $\tilde{e}(z) = \left(\cdots, \tilde{E}_x^{i,j,k}(z), \tilde{E}_y^{i,j,k}(z), \tilde{E}_z^{i,j,k}(z), \cdots\right)^T$ represents the transformed electric field $\mathbf{E}$ at discretized points of Yee cells shown in Figure 1 and $\tilde{h}(z), \tilde{j}(z), \widetilde{m}(z)$ represent $\mathbf{H}, \mathbf{J}, \mathbf{M}$ respectively following similar notations. $D_e(z) = \text{diag}\left(\cdots, \varepsilon_x^{i,j,k}(z), \varepsilon_y^{i,j,k}(z), \varepsilon_z^{i,j,k}(z), \cdots\right)$ is a diagonal matrix for permittivity, and $D_h(z) = \text{diag}\left(\cdots, \mu_x^{i,j,k}(z), \mu_y^{i,j,k}(z), \mu_z^{i,j,k}(z), \cdots\right)$ is a diagonal matrix for permeability. Both frequency-dependent permittivity and permeability are discretized into the z-transform space.[24] $\mathbf{C}_h, \mathbf{C}_e$ are the discretized curl operators. The resulted representation is equivalent to the one of the Finite-Differente Frequency-Domain (FDFD) method,[25] and is symmetric in an open region problem with lossy isotropic materials.

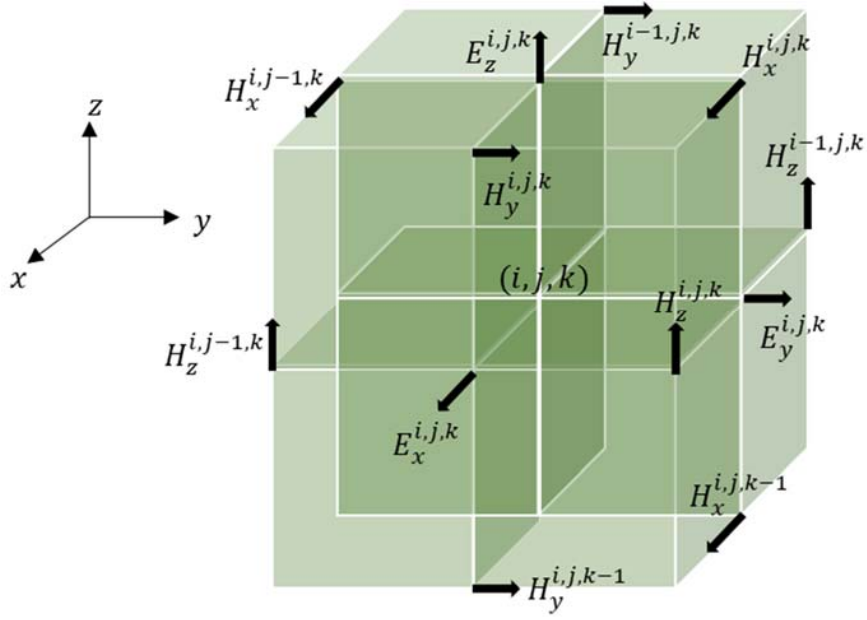

Figure 1: Component arrangement of the Yee cell. Each $E$ component is surrounded by four $H$ components and each $H$ component is surrounded by four $E$ components.

According to the adjoint method,[26] given an objective function $F\big(\mathbf{E}(\omega), \mathbf{H}(\omega)\big) \in \mathbb{R}$ which is numerically evaluated as $f\big(\tilde{e}(z), \tilde{h}(z)\big) \in \mathbb{R}$, variation of the numerical evaluation $f$ produced by variation of material properties $\delta D_e(z), \delta D_h(z)$ is given by

$$\delta f = \mathrm{Re}\left(-\begin{pmatrix}\tilde{e}_a(z)\\ \tilde{h}_a(z)\end{pmatrix}^T \delta \begin{pmatrix}-i\omega D_e(z) & \mathbf{C}_h\\ \mathbf{C}_e & i\omega D_h(z)\end{pmatrix}\begin{pmatrix}\tilde{e}(z)\\ \tilde{h}(z)\end{pmatrix}\right) \qquad (3)$$

$$= \mathrm{Re}\left(-\begin{pmatrix}\tilde{e}_a(z)\\ \tilde{h}_a(z)\end{pmatrix}^T \begin{pmatrix}-i\omega \delta D_e(z) & \\ & i\omega \delta D_h(z)\end{pmatrix}\begin{pmatrix}\tilde{e}(z)\\ \tilde{h}(z)\end{pmatrix}\right)$$

$$= \mathrm{Re}\left(i\omega \tilde{e}_a^T(z)\delta D_e(z)\tilde{e}(z) - i\omega \tilde{h}_a^T(z)\delta D_h(z)\tilde{h}(z)\right)$$

$$= \mathrm{Re}\left(\sum_{w,i,j,k} i\omega \tilde{E}_w^{i,j,k}(z)\tilde{E}_{a_w}^{\;i,j,k}(z)\delta \varepsilon_w^{i,j,k} - i\omega \tilde{H}_w^{i,j,k}(z)\tilde{H}_{a_w}^{\;i,j,k}(z)\delta \mu_w^{i,j,k}\right)$$

where $\left(\tilde{e}_a(z), \tilde{h}_a(z)\right)$ are the adjoint solution satisfying

$$\begin{pmatrix}-i\omega D_e(z) & \mathbf{C}_h\\ \mathbf{C}_e & i\omega D_h(z)\end{pmatrix}\begin{pmatrix}\tilde{e}_a(z)\\ \tilde{h}_a(z)\end{pmatrix} = \begin{pmatrix}\overline{\nabla_e f}\\ \overline{\nabla_h f}\end{pmatrix} \qquad (4)$$

The adjoint currents $\overline{\nabla_e f}$, $\overline{\nabla_h f}$ are conjugates of first derivatives of the objective function with respect to field variables[26]

$$\nabla_e f = \begin{pmatrix}\vdots\\ \partial f \,/\, \partial \tilde{E}_x^{i,j,k,n}\\ \partial f \,/\, \partial \tilde{E}_y^{i,j,k,n}\\ \partial f \,/\, \partial \tilde{E}_z^{i,j,k,n}\\ \vdots\end{pmatrix}, \nabla_h f = \begin{pmatrix}\vdots\\ \partial f \,/\, \partial \tilde{H}_x^{i,j,k,n}\\ \partial f \,/\, \partial \tilde{H}_y^{i,j,k,n}\\ \partial f \,/\, \partial \tilde{H}_z^{i,j,k,n}\\ \vdots\end{pmatrix} \qquad (5)$$

The symmetry of the system matrix suggests the same setup can be used for both forward simulation (2) and adjoint simulation (4), and the only difference between them is the source currents. In practice, open region problems are terminated with PMLs. The same sensitivity formula is applicable provided that PMLs are properly configured to absorb most outgoing waves. Additionally, the summation in (3) is the FDTD equivalence of integration and the Re operator comes from the fact that the objective function is real.

To solve the adjoint simulation, the objective function needs to be written explicitly with arguments from vectors $\tilde{e}(z)$ and $\tilde{h}(z)$ with respect to which the derivatives can be calculated to get the adjoint currents in frequncy domain. In practice, the objective function is calculated based on fields linearly interpolated from the original electromagnetic fields at Yee cells. The inverse DTFT is applied to convert frequency-domain adjoint currents to a time series such that the adjoint simulation(4) can be constructed and solved with FDTD in time domain. With solutions from the forward and the adjoint problems, sensitivity is evaluated according to (3). Details of implementation are provided in **Supporting Information 2**.

## 2.2 EXAMPLE: PLASMONIC SCATTERRER WITH VARYING PERMITTIVITY

We first demonstrate the accuracy of the discrete adjoint method by computing the sensitivity of the Poynting flux of a plasmonic scatterer with respect to a density parameter controlling its permittivity. To verify this sensitivity analysis, we benchmark it against the central differenec method. We calculate the Poynting flux over a specified region enclosing the plasmonic scatterer[27]

$$F = \int_S \text{Re}(\mathbf{E} \times \overline{\mathbf{H}}) \cdot \hat{n} dS \tag{6}$$

We consider a simple geometry for the plasmonic scatterer, a block with a size of $30 \times 30 \times 30$ nm³. The block has spatially uniform permittivity described by a density parameter $\rho$ using linear interpolation

$$\varepsilon(\rho) = \varepsilon_1 \rho + \varepsilon_2 (1 - \rho) \tag{7}$$

where $\varepsilon_1$ is permittivity of gold and $\varepsilon_2$ is permittivity of vacuum. According to (3), the adjoint sensitivity is given by

$$\frac{\partial f}{\partial \rho}(\rho) = \mathrm{Re}\left(\sum_{w,i,j,k} i\omega \tilde{E}_w^{i,j,k}(z) \tilde{E}_{a_w}^{i,j,k}(z)(\varepsilon_1 - \varepsilon_2)\right) \tag{8}$$

A plane wave of wavelength 800 nm is incident upon the block. The layout for the example is shown in Figure 2.

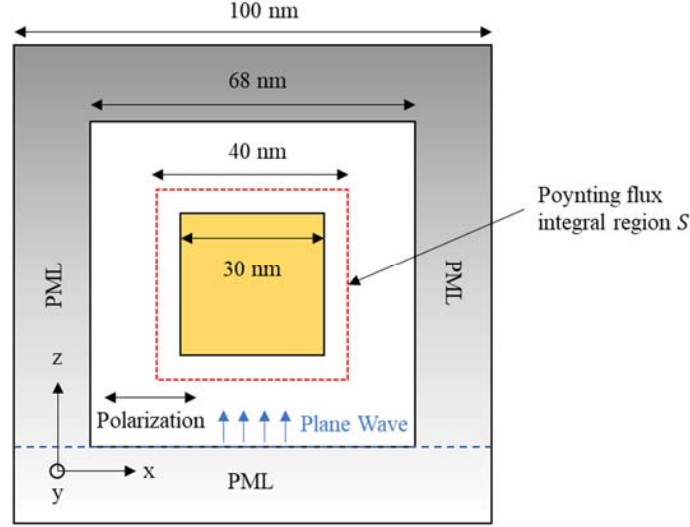

Figure 2: Layout for the example. A block of size $30 \times 30 \times 30$ nm$^3$ is seated in an open region with its axis aligned with coordinates. An incident plane wave travels in the $z^+$ direction. PMLs of 8-cell thickness is used to absorb any outgoing waves. The Poynting flux is integrated on the box surrounding the block.

We take the Lorentz-Drude model[28] as the dispersion model of gold. We linearly scale the strength of each susceptibility to implement the linearly interpolated permittivity (7). At 800 nm, the permittivity of gold is $\varepsilon_1 = \varepsilon_0(-22.3 - 2.03j)$ calculated from the Lorentz-Drude model. We use a 2 nm Yee cell length to run forward simulations to calculate the Poynting flux versus the density parameter $\rho$ and run adjoint simulations to get the sensitivity with respect to $\rho$. Result is shown in Figure 3(a).

We then compare the adjoint sensitivity (8) with sensitivity calculated from central difference scheme

$$\frac{\partial f}{\partial \rho}(\rho) \cong \frac{f(\rho + \Delta\rho) - f(\rho - \Delta\rho)}{2\Delta\rho} \tag{9}$$

to verify the accuracy of the discrete adjoint sensitivity formula. Figure 3(b-d) compare the two in zoomed-in ranges of $\rho$ with $\Delta\rho = 0.001$. The obvious agreement between the two confirms the validity of the discrete adjoint sensitivity formula.

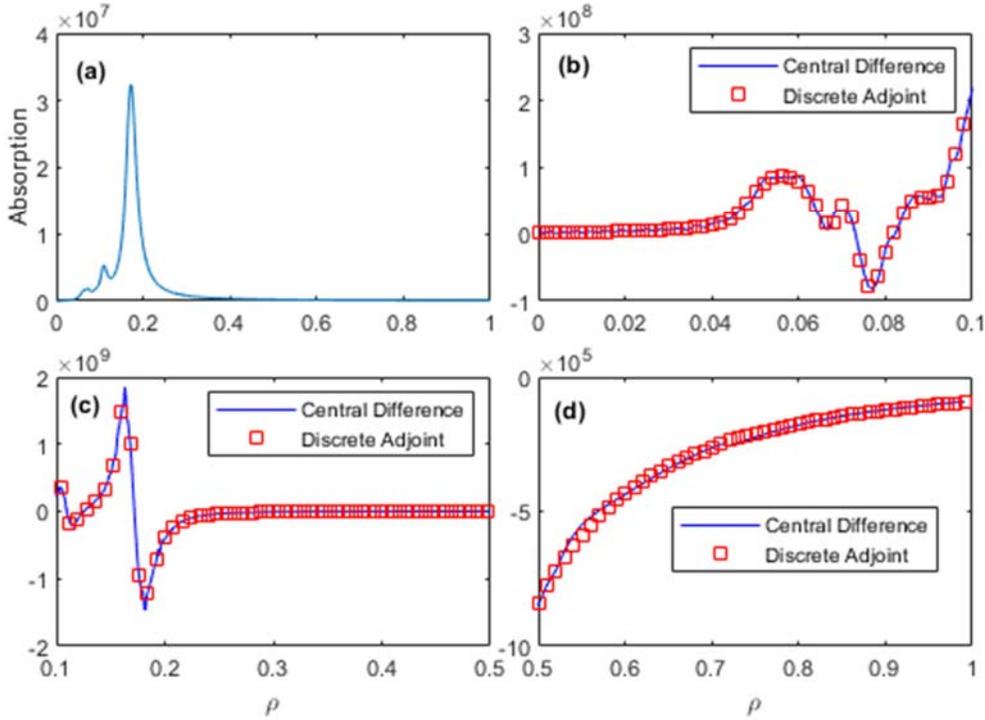

Figure 3: (a): The Poynting flux $F$ of the block scatterer vs the density parameter $\rho$. (b)-(d): Comparison of the central difference sensitivity and the discrete adjoint sensitivity. (b): $\rho \in [0, 0.1]$. (c): $\rho \in [0.1, 0.5]$. (d): $\rho \in [0.5, 1]$.

## 3. DENSITY-BASED TOPOLOGY OPTIMIZATION

The adjoint method calculates gradients with only two simulations, the convenience of which makes gradient-based optimizations that involve millions of variables possible, e.g., permittivity at each discretized location. The density-based topology optimization is one of many gradient-based optimizations that is extensively used in inverse design of photonic devices.[8–10,29–35] It represents the design using a vector $\rho$ of continuous density parameters $\rho^i \in [0,1]$ that are mapped to the physical material distribution by a material interpolation scheme and utilizes the adjoint method to obtain gradients. For example, we can assign value to a permittivity $\varepsilon^i(z)$ in FDTD using linear interpolation

$$\varepsilon^i(\rho^i, z) = \varepsilon_1(z)\rho^i + \varepsilon_2(z)(1 - \rho^i) \tag{10}$$

where $\varepsilon_1$ is the permittivity for the material domain and $\varepsilon_2$ is the permittivity for the void domain (usually $\varepsilon_2 = \varepsilon_0$). In our approach, a rectangular design domain is constructed with a regular grid (Cartesian lattice). Each point in the grid is assigned a density parameter $\rho^i$ and linear interpolation is used to map grid density parameters into Yee lattice density parameters, which then determines the material property by the linear material interpolation.

It is also demonstrated[9] that in topology optimization of plasmonics the linear interpolation in (10) can cause non-physical field amplification. Such amplification is unique for problems that involve negative permittivity, e.g. plasmonic materials, and destroys the stable convergence of the optimization. A non-linear material interpolation scheme is introduced[9] to improve the convergence behavior and hence incorporated into our method. In the non-linear method, the refractive index is linearly interpolated:

$$\varepsilon^i(\rho^i, z) = \left(\rho^i\sqrt{\varepsilon_2(z)} + (1 - \rho^i)\sqrt{\varepsilon_1(z)}\right)^2 \tag{11}$$

$$= \rho^{i^2}\varepsilon_2(z) + (1 - \rho^i)^2\varepsilon_1(z) + 2\rho^i(1 - \rho^i)\sqrt{\varepsilon_1(z)\varepsilon_2(z)}$$

In contrast to FEM, there is no direct way of implementing the non-linear interpolation in FDTD since $\sqrt{\varepsilon_1(z)\varepsilon_2(z)}$ is not a rational function in general. However, if only one frequency is concerned in the optimization (which is true for many plasmonic applications), we can replace $\sqrt{\varepsilon_1(z)\varepsilon_2(z)}$ with a rational function $\varepsilon_3(z)$ that has the same value at this frequency. **Supporting Information 3** details the implementation of non-linear material interpolation in FDTD for one frequency.

The three-field density representation[36] is adopted as the regularization technique in our method to remove very small features to ensure manufacturability, which uses a density filter followed by a projection. Each density value $\rho^i$ is reassigned as a weighted sum of its neighbors:

$$\hat{\rho}^i = \frac{\sum_{j \in N_i} w_{ij} \rho^j}{\sum_{j \in N_i} w_{ij}} \tag{12}$$

We use a Gaussian distribution function[37] for the filter function $w_{ij}$

$$w_{ij} = e^{-0.5\left(\frac{\|x_i - x_j\|}{R/2}\right)^2} \tag{13}$$

where $R$ denotes the filter radius which is used to truncate the filter and is deemed as the minimum feature size of the design. Density filters can effectively remove small features during the optimization process but can end up creating more gray transition regions. Therefore, a projection operator is used to project the filtered density to 0/1. The following projection operator[38] is used in our method:

$$\bar{\hat{\rho}}^i = \frac{\tanh(\beta\eta) + \tanh\left(\beta\left(\hat{\rho}^i - \eta\right)\right)}{\tanh(\beta\eta) + \tanh\left(\beta(1 - \eta)\right)} \tag{14}$$

where $\beta$ is a parameter used to control the strength of the projection and $\eta \in [0,1]$ defines the projection level. Eventually, projected density parameters $\bar{\hat{\rho}}^i$ are used as the actual density in the simulation. To prevent the optimization from getting stuck in local minimum earlier in the process due to the density being projected to 0/1 immediately, projection strength $\beta$ is increased after the objective function has not changed much after a certain number of iterations, $n_{sc}$, mimicking the $\beta$-continuation scheme.[38]

We programmed a FDTD solver in C++ for handling solutions of forward and adjoint problems. The solver allows specifying material property of each Yee cell point so that we are able implement the non-linear material interpolation. A Python interface was designed to communicate to the solver, calculate sensitivities, and evolve solutions in each inner iteration using the method of moving asymptotes (MMA).[39] The overall flow diagram for the density-based topology optimization with the discrete adjoint sensitivity formula is shown in Figure 4. The $\beta$-continuation scheme consists of an inner loop that evolves inner design density $\rho_i$ and an outer loop that changes projection level. It should be noted that in the step that involves MMA one or more forward simulations are run until an approximating subproblem is found to improve the objective function.[40] In addition, only one adjoint simulation is run per iteration in the inner loop. Therefore, the number of forward simulations is usually greater than the number of adjoint simulations.

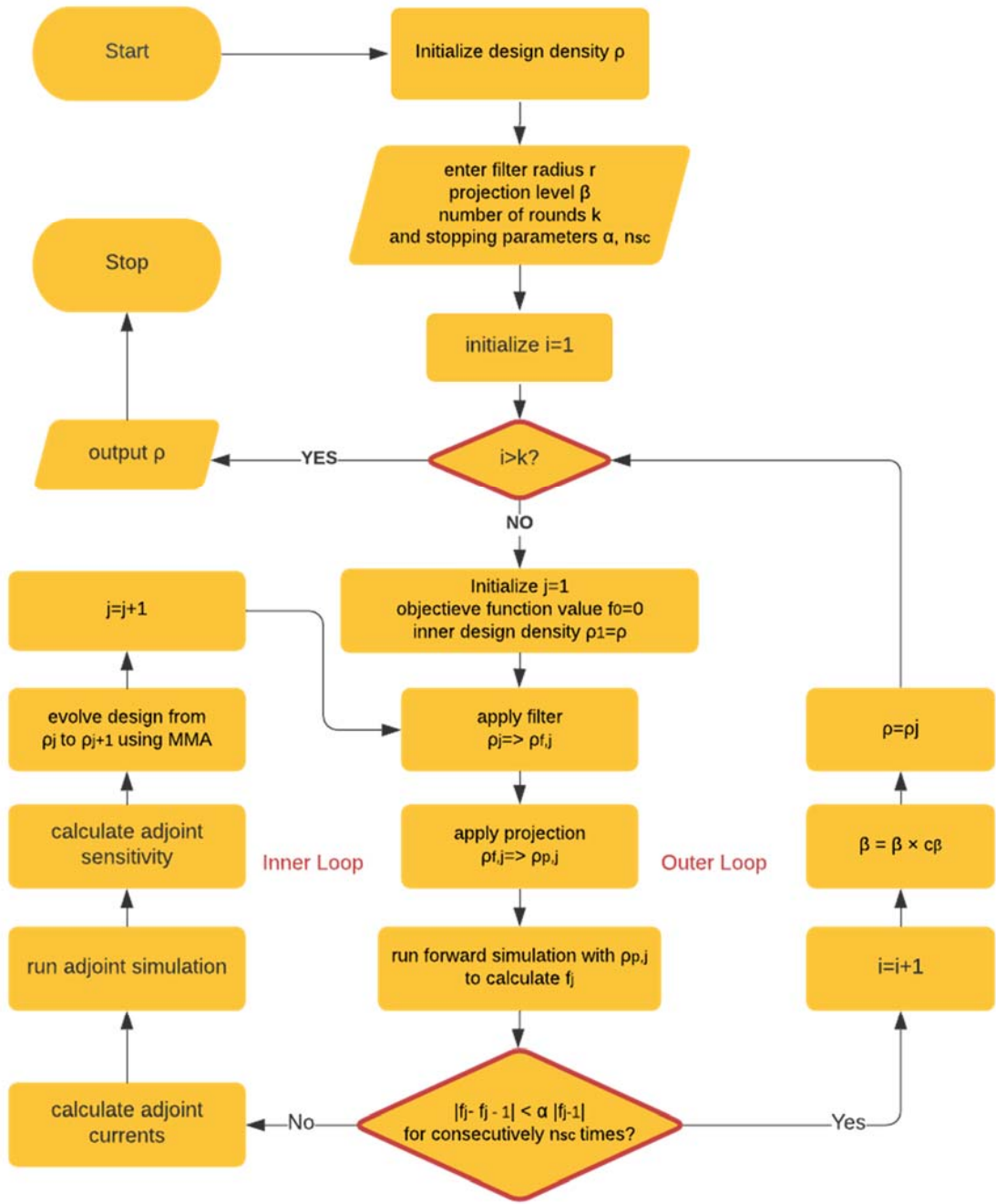

Figure 4: Flow diagram of density-based topology optimization. The diagram exemplifies an optimization that runs $k$ rounds where $c_\beta$ is rate of change for projection level between each iteration in the outer loop.

# 4. CASE STUDY: RECONSTRUCTION OF ELECTRIC FIELDS OF A PLASMONIC BOWTIE APERTURE

Plasmonic bowtie apertures are known to produce highly localized fields and have potential applications including optical lithography[4] and high density data storage.[41] The enhanced electric fields are confined within only a region of the nanometer length scale near the surface of the nanostructures and decay significantly thereafter.[42] In this example, reconstruction of the electric fields of a plasmonic bowtie aperture is carried out to exemplify inverse design with a non-linear frequency-domain objective function. The reconstruction is realized by minimizing an objective function that measures the difference between the design electric field $\mathbf{E}$ and the objective electric field $\mathbf{E}_0$ produced by the plasmonic bowtie aperture. In other words, the objective function is defined as the normalized difference:

$$F(\mathbf{E}) = \int_S (\|\mathbf{E}\| - \|\mathbf{E}_0\|)^2 \, dS \Big/ \int_S \|\mathbf{E}_0\|^2 \, dS \qquad (15)$$

where $S$ is chosen as the exit plane of the bowtie structure. The layout for the optimization and target geometry are shown in Figure 5. The Yee cell length is 2 nm to accurately capture the near field.[43] The optimization follows the flow diagram in Figure 4 and uses a $N_x \times N_y \times N_z$ rectangular grid for design density. Table 1 lists parameters used in this optimization. We enforce the density to be constant along the z direction so that the optimization produces a planar structure. The optimization algorithm starts with $\rho = 1$ and runs for 5 iterations in the outer loop, which constitutes a total of 50 adjoint simulations and 98 forward simulations.

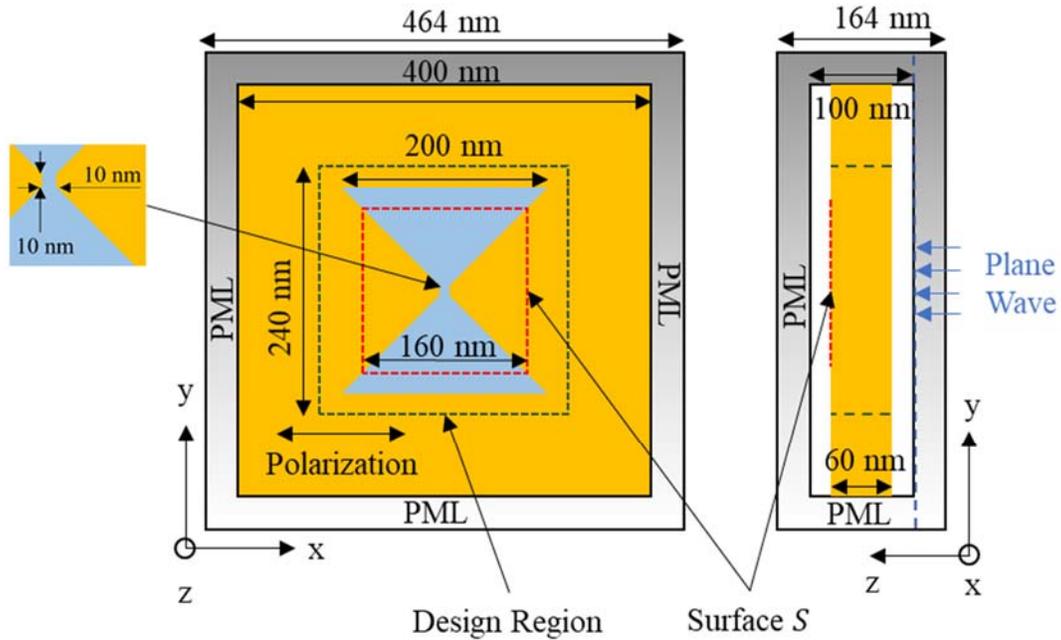

Figure 5: Layout for the optimization. All rectangles are squares with their centers alighned. An incident wave of wavelength 800 nm is traveling in direction perpendicular to the gold layer marked in yellow.

Table 1 Parameters used in optimization. Note that $N_z$ discretized points are equivalent to $N_z - 1$ discretized intervals.

| Parameter [unit] | Value |
|---|---|
| $R$ [nm] | 8 |
| $\beta$ | 1 |
| $k$ | 5 |
| $\alpha$ | 0.01 |
| $c_r$ | 1.0 |
| $c_\beta$ | 1.8 |
| $N_x, N_y, N_z$ | 121, 121, 31 |
| $n_{sc}$ | 3 |

The objective function in the final design is evaluated as 0.0076, suggesting a successful recovery of the electric field on the surface $S$ as shown in Figure 6. Although symmetry is not enforced during the optimization, the density distributions are symmetric with the same bowtie structure within the surface $S$ region. The features at the corners of the final design are artifacts of the optimization and do not influence the centered fields as they are almost identical for the original and optimization generated designs shown in Figure 6c and Figure 6d.

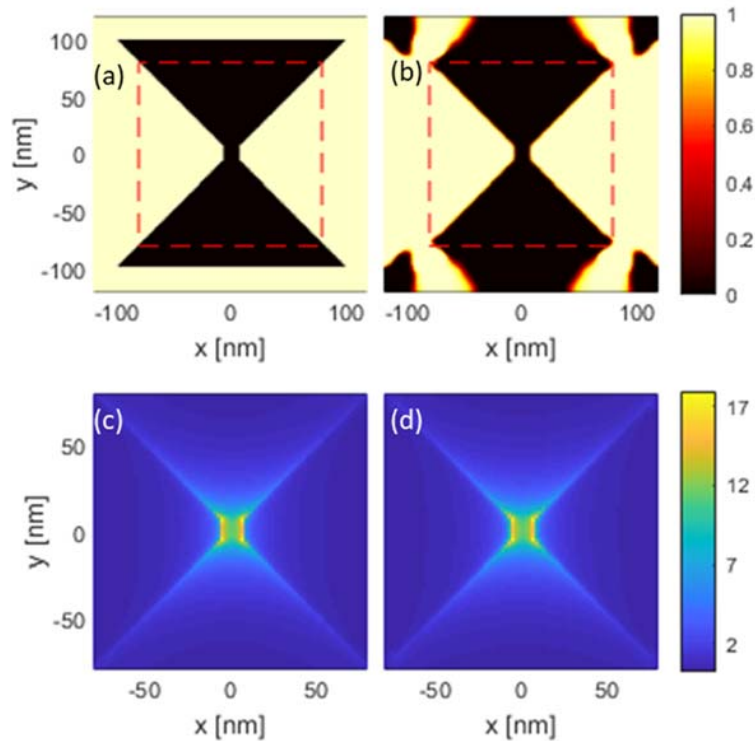

Figure 6: Density distribution of the original bowtie structure (a) and the final design structure (b) overlayed with surface $S$ marked by red dashed line and their respective electric fields distribution on surface $S$ (c,d). The electric fields distribution is normalized by incident wave intensity.

Figure 7 shows intermediate iterations during the optimization. The sharp decreases in the objective function happen at the beginning of each iteration in the outer loop when the projection level $\beta$ increases. Higher value of $\beta$ brings in higher contrast of air and gold to the boundary, which contributes to the generation of highly localized fields. During the early iterations, the general shape of the final design already emerges. Early emergence of a clearly defined shape has also been reported[9,42] using density-based topology optimization.

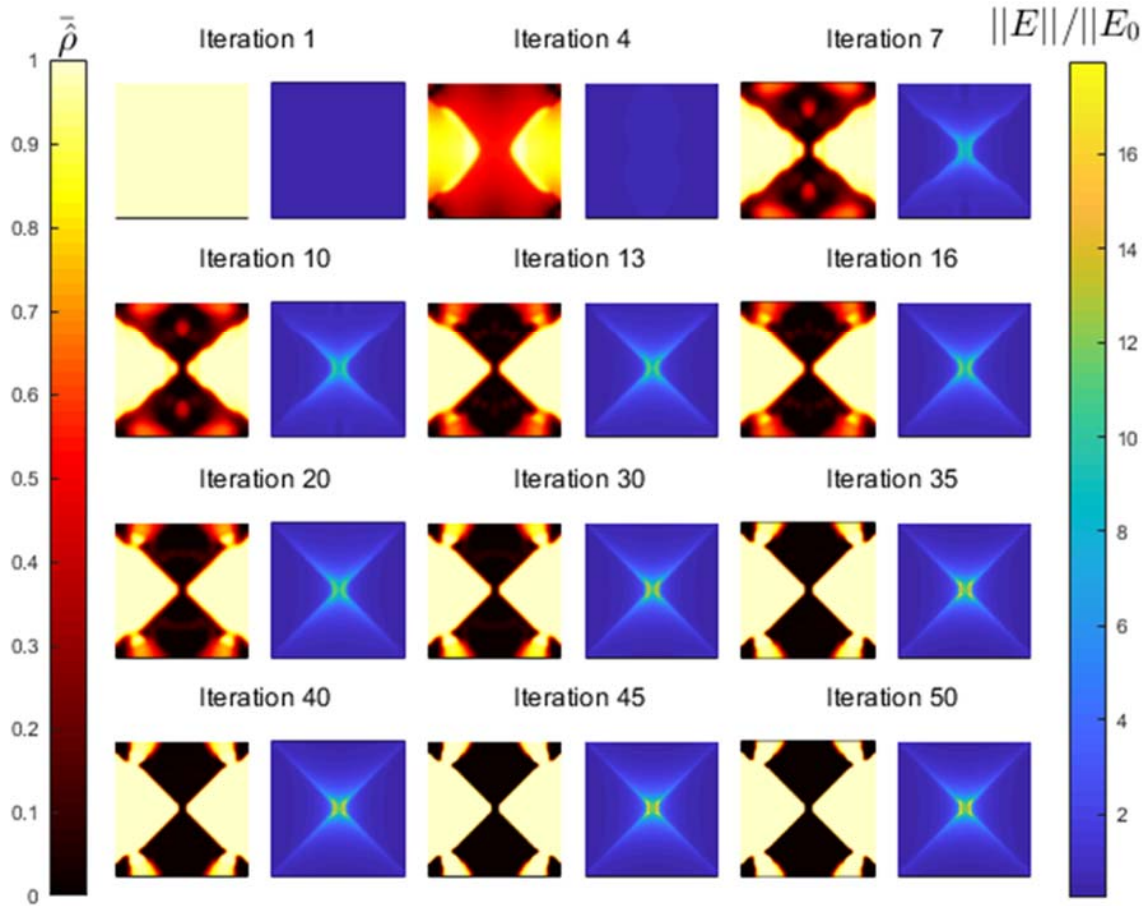

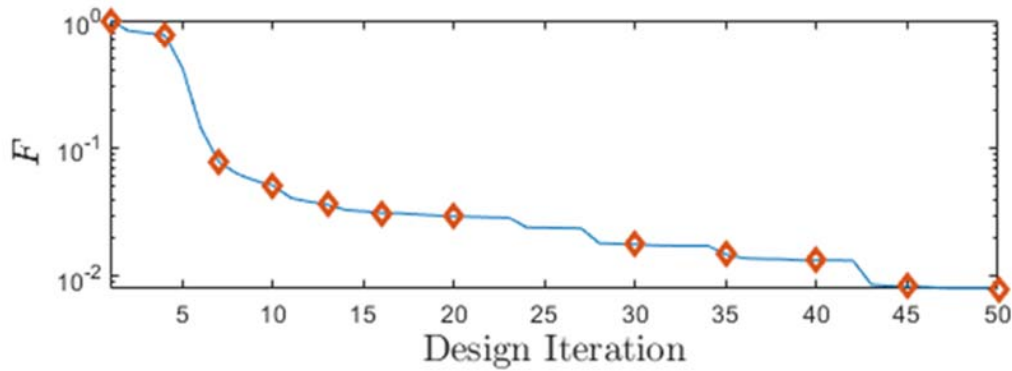

Figure 7: Distributions of the projected density $\bar{\bar{\rho}}$ (left) and the field magnitude (right) in some intermediate steps showing evolution of the density. Each iteration is marked with diamond in the objective function graph at the bottom. The objective function is normalized so that it is evaluated as 1 in the 1st iteration.

## 5. CONCLUSION

We present a method for carrying out inverse design of plasmonic structures using FDTD that incorporates the discrete adjoint method in frequency domain. This method can be built upon any existing FDTD program and accepts general differentiable objective functions. We provide detailed derivations of the adjoint method and validate it with a numerical example involving a plasmonic scatterer. The adjoint sensitivities match very well with sensitivities obtained with the central-difference scheme. Based on the discrete adjoint method, we develop a density-based topology optimization framework and improve the robustness by using the filtering-and-projection regularization and the non-linear material interpolation. For illustration, we use the method to successfully reconstruct electric fields of a bowtie plasmonic structure.

## ACKNOWLEDGEMENTS

Support to this work by ASTC – the Advanced Storage Technology Consortium and the National Science Foundation is acknowledged.

**Supporting Information: Inverse Design of Plasmonic Structures with FDTD**

Zhou Zeng and Xianfan Xu*

School of Mechanical Engineering and Birck Nanotechnology Center

Purdue University, West Lafayette, IN 47906

* Corresponding author.

Email addresses: zeng133@purdue.edu (Zhou Zeng), xxu@ecn.purdue.edu (Xianfan Xu)

## 1. FDTD IN FREQUENCY DOMAIN

In the main text, we stated that the frequency-domain matrix representation of FDTD is symmetric in an open region problem with lossy isotropic materials. Here we provide details of the derivation.

We start with the basic formulation of three dimensional FDTD method with lossy isotropic materials. For convenience, we index each field component by the position of the corresponding Yee cell,[1] which is illustrated below.

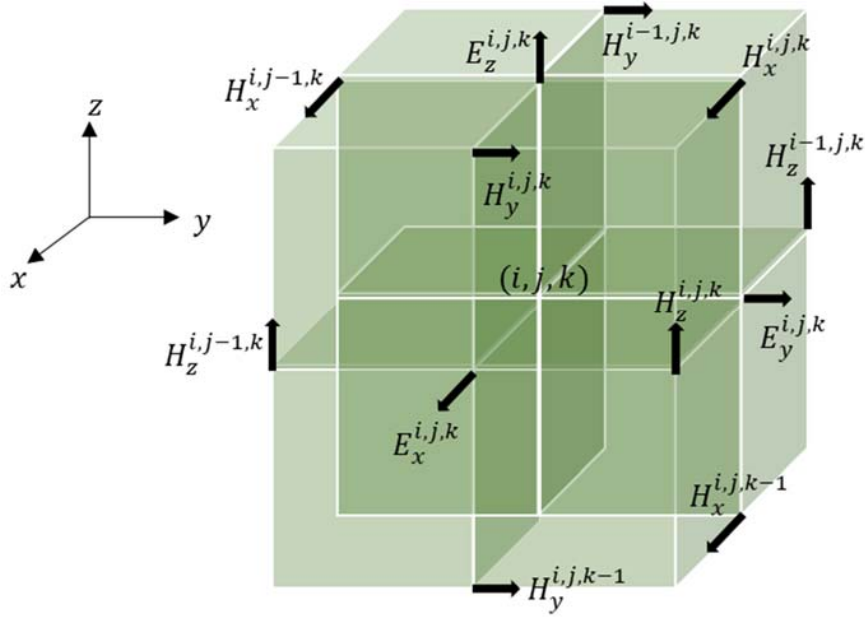

Figure 8. Component arrangement of the Yee cell. Each $E$ component is surrounded by four $H$ components and each $H$ component is surrounded by four $E$ components.

With the leapfrog time-stepping scheme, we can write down evaluations of all field components

$$E_x^{i,j,k,n} = E\big((i + 0.5)\Delta x, j\Delta y, k\Delta z, n\Delta t\big) \tag{16a}$$

$$E_y^{i,j,k,n} = E(i\Delta x, (j + 0.5)\Delta y, k\Delta z, n\Delta t) \tag{16b}$$

$$E_z^{i,j,k,n} = E(i\Delta x, j\Delta y, (k + 0.5)\Delta z, n\Delta t) \tag{16c}$$

$$H_x^{i,j,k,n} = E(i\Delta x, (j + 0.5)\Delta y, (k + 0.5)\Delta z, (n + 0.5)\Delta t) \tag{16d}$$

$$H_y^{i,j,k,n} = E\big((i + 0.5)\Delta x, j\Delta y, (k + 0.5)\Delta z, (n + 0.5)\Delta t\big) \tag{16e}$$

$$H_z^{i,j,k,n} = E\big((i + 0.5)\Delta x, (j + 0.5)\Delta y, k\Delta z, (n + 0.5)\Delta t\big) \tag{16f}$$

where $\Delta x, \Delta y, \Delta z$ are space steps and $\Delta t$ is the time step. The central-difference expressions for the

space and time derivatives convert the two differential curl equations

$$\frac{\partial \mathbf{B}}{\partial t} = -\nabla \times \mathbf{E} - \mathbf{M} \tag{17}$$

$$\frac{\partial \mathbf{D}}{\partial t} = \nabla \times \mathbf{H} - \mathbf{J} \tag{18}$$

into two finite difference equations

$$\frac{b^n - b^{n-1}}{\Delta t} = -\mathbf{C}_e e^n - m^n \tag{19}$$

$$\frac{d^{n+1} - d^n}{\Delta t} = \mathbf{C}_h h^n - j^n \tag{20}$$

where

$$d^n = \begin{pmatrix} \vdots \\ D_x^{i,j,k,n} \\ D_y^{i,j,k,n} \\ D_z^{i,j,k,n} \\ \vdots \end{pmatrix}, h^n = \begin{pmatrix} \vdots \\ H_x^{i,j,k,n} \\ H_y^{i,j,k,n} \\ H_z^{i,j,k,n} \\ \vdots \end{pmatrix}, j^n = \begin{pmatrix} \vdots \\ J_x^{i,j,k,n} \\ J_y^{i,j,k,n} \\ J_z^{i,j,k,n} \\ \vdots \end{pmatrix} \tag{21}$$

and

$$b^n = \begin{pmatrix} \vdots \\ B_x^{i,j,k,n} \\ B_y^{i,j,k,n} \\ B_z^{i,j,k,n} \\ \vdots \end{pmatrix}, e^n = \begin{pmatrix} \vdots \\ E_x^{i,j,k,n} \\ E_y^{i,j,k,n} \\ E_z^{i,j,k,n} \\ \vdots \end{pmatrix}, m^n = \begin{pmatrix} \vdots \\ M_x^{i,j,k,n} \\ M_y^{i,j,k,n} \\ M_z^{i,j,k,n} \\ \vdots \end{pmatrix} \tag{22}$$

are column vectors representing the relevant fields; $\mathbf{C}_h$, whose nonzero elements are $\pm 1/\Delta x$, $\pm 1/\Delta y$ and $\pm 1/\Delta z$, is a matrix for the curl operator on the $H$ components while $\mathbf{C}_e$ is a matrix for the curl operator on the $E$ components. The Discrete Time Fourier Transform (DTFT)[2] converts a discrete-time signal, which is a sequence of real numbers, into a complex frequency-domain representation given below:

$$\tilde{x}(z) = \sum_{n=0}^{\infty} x[n] z^{-n} \; z = e^{i\omega \Delta t} \tag{23}$$

where $z = e^{i\omega\Delta t}$ is used for simplicity and tilde denotes DTFT of the corresponding variable. Applying DTFT to the two finite difference equations (19), (20) yields

$$\frac{z\tilde{d}(z) - \tilde{d}(z)}{\Delta t} = \mathbf{C}_h \tilde{h}(z) - \tilde{j}(z) \tag{24}$$

$$\frac{\tilde{b}(z) - z^{-1}\tilde{b}(z)}{\Delta t} = -\mathbf{C}_e \tilde{e}(z) - \widetilde{m}(z) \tag{25}$$

The generalization of FDTD algorithms of linear dispersions[3] is given below

$$\tilde{d}(z) = D_e(z)\tilde{e}(z) \tag{26}$$

$$\tilde{b}(z) = D_h(z)\tilde{h}(z) \tag{27}$$

where $D_e(z) = \text{diag}\left(\cdots, \varepsilon_x^{i,j,k}(z), \varepsilon_y^{i,j,k}(z), \varepsilon_z^{i,j,k}(z), \cdots\right)$ is a diagonal matrix for permittivity, and $D_h(z) = \text{diag}\left(\cdots, \mu_x^{i,j,k}(z), \mu_y^{i,j,k}(z), \mu_z^{i,j,k}(z), \cdots\right)$ is a diagonal matrix for permeability. $\omega\Delta t$ is often very small to resolve aliasing in DTFT such that the following approximations can be made

$$z \cong 1 + i\omega\Delta t \tag{28}$$

$$z^{-1} \cong 1 - i\omega\Delta t \tag{29}$$

Combining the two difference equations (24), (25), the DTFT of constitutive equations (26), (27) and the two approximations (28),(29), we have

$$\begin{pmatrix} -i\omega D_e(z) & \mathbf{C}_h \\ \mathbf{C}_e & i\omega D_h(z) \end{pmatrix} \begin{pmatrix} \tilde{e}(z) \\ \tilde{h}(z) \end{pmatrix} = \begin{pmatrix} \tilde{j}(z) \\ -\widetilde{m}(z) \end{pmatrix} \tag{30}$$

which is equivalent to the Finite-Difference Frequency-Domain (FDFD) formulation.[4] We claim that for open region problems with isotropic materials, the operator in the above equation is symmetric. Obviously, $D_e(z)$ and $D_h(z)$ are symmetric since they are diagonal matrices. Therefore, we only need to show $\mathbf{C}_e^T = \mathbf{C}_h$. We define $\mathbf{C}_h\left(\tilde{E}_w^{i,j,k}, \tilde{H}_{w_1}^{i_1,j_1,k_1}\right)$ as the entry of matrix $\mathbf{C}_h$ that appears in the linear equation extracted from Equation (2)

$$-i\omega\varepsilon_x^{i,j,k}\tilde{E}_x^{i,j,k} + \sum_{w_1,i_1,j_1,k_1} \mathbf{C}_h\big(\tilde{E}_w^{i,j,k},\tilde{H}_{w_1}^{i_1,j_1,k_1}\big)\tilde{H}_{w_1}^{i_1,j_1,k_1} = \tilde{J}_x^{i,j,k} \tag{31}$$

Likewise, the definition of $\mathbf{C}_e\big(\tilde{H}_w^{i,j,k},\tilde{E}_{w_1}^{i_1,j_1,k_1}\big)$ follows

$$i\omega\mu_y^{i,j,k}H_y^{i,j,k} + \sum_{w_1,i_1,j_1,k_1} \mathbf{C}_e\big(\tilde{H}_w^{i,j,k},\tilde{E}_{w_1}^{i_1,j_1,k_1}\big)E_{w_1}^{i_1,j_1,k_1} = -\tilde{M}_x^{i,j,k} \tag{32}$$

We write down the complete finite-difference equations Yee's algorithm[1]

$$-i\omega\varepsilon_x^{i,j,k}\tilde{E}_x^{i,j,k} + \frac{\tilde{H}_z^{i,j,k}-\tilde{H}_z^{i,j-1,k}}{\Delta y} - \frac{\tilde{H}_y^{i,j,k}-\tilde{H}_y^{i,j,k-1}}{\Delta z} = \tilde{J}_x^{i,j,k} \tag{33a}$$

$$-i\omega\varepsilon_y^{i,j,k}\tilde{E}_y^{i,j,k} + \frac{\tilde{H}_x^{i,j,k}-\tilde{H}_x^{i,j,k-1}}{\Delta z} - \frac{\tilde{H}_z^{i,j,k}-\tilde{H}_z^{i-1,j,k}}{\Delta x} = \tilde{J}_y^{i,j,k} \tag{33b}$$

$$-i\omega\varepsilon_z^{i,j,k}\tilde{E}_z^{i,j,k} + \frac{\tilde{H}_y^{i,j,k}-\tilde{H}_y^{i-1,j,k}}{\Delta x} - \frac{\tilde{H}_x^{i,j,k}-\tilde{H}_x^{i,j-1,k}}{\Delta y} = \tilde{J}_z^{i,j,k} \tag{33c}$$

$$-\frac{\tilde{E}_y^{i,j,k+1}-\tilde{E}_y^{i,j,k}}{\Delta z} + \frac{\tilde{E}_z^{i,j+1,k}-\tilde{E}_z^{i,j,k}}{\Delta y} + i\omega\mu_x^{i,j,k}\tilde{H}_x^{i,j,k} = -\tilde{M}_x^{i,j,k} \tag{33d}$$

$$-\frac{\tilde{E}_z^{i+1,j,k}-\tilde{E}_z^{i,j,k}}{\Delta x} + \frac{\tilde{E}_x^{i,j,k+1}-\tilde{E}_x^{i,j,k}}{\Delta z} + i\omega\mu_y^{i,j,k}\tilde{H}_y^{i,j,k} = -\tilde{M}_y^{i,j,k} \tag{33e}$$

$$-\frac{\tilde{E}_x^{i,j+1,k}-\tilde{E}_x^{i,j,k}}{\Delta y} + \frac{\tilde{E}_y^{i+1,j,k}-\tilde{E}_y^{i,j,k}}{\Delta x} + i\omega\mu_y^{i,j,k}\tilde{H}_y^{i,j,k} = -\tilde{M}_z^{i,j,k} \tag{33f}$$

Then the conclusion $\mathbf{C}_h\big(\tilde{E}_w^{i,j,k},\tilde{H}_{w_1}^{i_1,j_1,k_1}\big) = \mathbf{C}_e\big(\tilde{H}_{w_1}^{i_1,j_1,k_1},\tilde{E}_w^{i,j,k}\big)$ is immediate. As an example of symmetry, the following is satisfied

$$\frac{1}{\Delta y} = \mathbf{C}_e\big(E_z^{i,j,k},\tilde{H}_x^{i,j-1,k}\big) = \mathbf{C}_h\big(\tilde{H}_x^{i,j,k},E_z^{i,j+1,k}\big) = \mathbf{C}_h\big(\tilde{H}_x^{i,j-1,k},E_z^{i,j,k}\big) \tag{34}$$

## 2. IMPLEMENTATION OF THE ADJOINT METHOD IN FDTD

In the main text, we discussed the adjoint method and the sensitivity formula. Here we provide implementation details.

First to solve the adjoint simulation, the objective function needs to be written explicitly with arguments from vectors $\tilde{e}(z)$ and $\tilde{h}(z)$ with respect to which the derivatives can be calculated. In practice, the objective function is always calculated based on fields interpolated from the original electromagnetic fields at the Yee cell. To simplify notation, we assume that the objective function only depends on electric field $\tilde{e}(z)$, which is the case for most problems. Denote the interpolated electric fields by

$$\hat{e}(z) = \begin{pmatrix} \hat{\tilde{E}}_1(z) \\ \hat{\tilde{E}}_2(z) \\ \vdots \\ \hat{\tilde{E}}_n(z) \end{pmatrix} \tag{35}$$

Assuming linear interpolation, each interpolated electric field is given by

$$\hat{\tilde{E}}_i = \sum_{j \in N_i} w_{ij} \tilde{E}_j \tag{36}$$

where $N_i$ is the neighboring element for each interpolated electric field and $w_{ij}$ is the linear interpolation weight for the ith point from its j[th] neighbor.[5] We discard the original index system $\tilde{E}_x^{i,j,k}$ to avoid cluttered scripts. By the chain rule, we have derivatives of objective function with respect to original electric fields at the Yee cell given by

$$\frac{\partial f}{\partial \tilde{E}_i(z)} = \sum_{i \in N_j} \frac{\partial f}{\partial \hat{\tilde{E}}_j(z)} w_{ji} \tag{37}$$

where $\frac{\partial f}{\partial \tilde{E}_j(z)}$ on the right-hand side can be given an analytic form depending on the user-defined formula of objective function. Mathematically, the above equation exhibits a well-known conecpt originating in multi-grid methods – the transpose of interpolation.[6]

Finally to evaluate gradients at discretized points of the design region, we also apply the chain rule. Assuming linear interpolation is used to map design density parameters into Yee lattice density parameters, we can also apply transpose of interpolation to map gradients at Yee lattice back to discretized points in the design region. Other non-linear interpolation schemes can be applied, but require much more sophisticated implementation of chain rule.

## 3.  NON-LINEAR MATERIAL INTERPOLATION IN FDTD FOR ONE FREQUENCY

In the main text, we presented the non-linear material interpolation. Here we give details of implementation.

The refractive index is linear interpolated in the non-linear material method[7]

$$\varepsilon(\rho, \omega) = \left(\rho\sqrt{\varepsilon_2(\omega)} + (1-\rho)\sqrt{\varepsilon_1(\omega)}\right)^2 \tag{38}$$

$$= \rho^2\varepsilon_2(\omega) + (1-\rho)^2\varepsilon_1(\omega) + 2\rho(1-\rho)\sqrt{\varepsilon_1(\omega)\varepsilon_2(\omega)}$$

There is no direct way of implementing the non-linear interpolation in FDTD since $\sqrt{\varepsilon_1(\omega)\varepsilon_2(\omega)}$ is not a rational function in general. Therefore, we replace $\sqrt{\varepsilon_1(\omega)\varepsilon_2(\omega)}$ with a rational function $\varepsilon_3(\omega)$ that has the same value for the single frequency of interest. In our method, we use a single

Lorentz pole[8] to construct $\varepsilon_3(\omega)$. Denote the value for $\sqrt{\varepsilon_1(\omega)\varepsilon_2(\omega)}$ at frequency $\omega_0$ as $\varepsilon_{no}$ and define $\varepsilon_3(\omega)$ as below

$$\varepsilon_3(\omega) = \text{Re}(\varepsilon_{no}) + \text{Im}(\varepsilon_{no}) \frac{\omega_0^2}{\omega_0^2 - \omega^2 - i\omega\omega_0} \tag{39}$$

Immediately we have $\varepsilon_3(\omega_0) = \varepsilon_{no}$. As a result, the material model becomes

$$\varepsilon'(\rho, \omega) = \rho^2 \varepsilon_2(\omega) + (1-\rho)^2 \varepsilon_1(\omega) + 2\rho(1-\rho)\varepsilon_3(\omega) \tag{40}$$

Using the Lorentz-Drude[8] material model for $\varepsilon_1(\omega)$ and $\varepsilon_2(\omega)$, Equation (40) is essentially another Lorentz-Drude model with $\rho$ parameterized instantaneous dielectric constants and $\rho$ parameterized intensities for susceptibilities

$$\varepsilon'(\rho, \omega) = \varepsilon_\infty(\rho) + \sum_{n=1}^{n=n_0} \sigma_n(\rho) \frac{\omega_n^2}{\omega_n^2 - \omega^2 - i\omega\gamma_n} \tag{41}$$

In the inverse design, each discretized point in the Yee lattice is assigned a density parameter $\rho$. If we store the complete material model from Equation (41) for each discretized point and use an array of structs to hold the information, we would end up bloating memory and decreasing data locality[9] at the same time. The program would be both memory consuming and slow. Noticing that the frequency-dependent terms in the material model are common among all discretized points, we can store all of the frequency-dependent terms into a globally accessible array of structs and only keep the information of intensities at the discretized point. In other words, we store $\varepsilon_\infty(\rho)$ and $\sigma_n(\rho)$ for each discretized point into an array of structs and store frequency-dependent terms $\frac{\omega_n^2}{\omega_n^2 - \omega^2 - i\omega\gamma_n}$ separately into another array of structs.